\documentclass[prl,twocolumn]{revtex4}

\usepackage{mathrsfs,amsmath,graphicx}

\def\scr{\mathscr}

 \def\SV{{\scr V}}

\def\Bid{{\mathchoice {\rm {1\mskip-4.5mu l}} {\rm
{1\mskip-4.5mu l}} {\rm {1\mskip-3.8mu l}} {\rm {1\mskip-4.3mu l}}}}

\def\avg#1{\langle#1\rangle}    \def\<{\langle}         \def\>{\rangle}

\def\sig{\sigma}        \def\del{\delta}        \def\Del{\Delta}
\def\eps{\epsilon}

  \def\V0{{\mathbf 0}}

  \def\B0{{\mathbf 0}}
\def\Br{{\bf r}}   
\def\Bj{{\mathbf j}}
\def\Bk{{\mathbf k}}  \def\Bq{{\mathbf q}}

  \def\BH{{\mathbf H}}

\def\td#1{{\tilde{#1}}}

\def\be{\begin{equation}}       \def\ee{\end{equation}}
\def\bea{\begin{eqnarray}}      \def\eea{\end{eqnarray}}

\begin{document} 

\title{Anomalous quantum mass flow of atoms in $p$-wave resonance
}

\author{W. Vincent Liu}
\affiliation{Department of Physics and Astronomy, University of
  Pittsburgh, Pittsburgh, PA 15260, USA}


\begin{abstract}
  
  I analyze an atomic Fermi gas with a planar $p$-wave interaction,
  motivated by the experimentally observed anisotropy in $p$-wave
  Feshbach resonances.  An axial superfluid state is verified.  A
  domain wall object is discovered to be a new topological defect of
  this superfluid and an explicit solution has been found.  Gapless
  quasiparticles appear as bound states on the wall, dispersing in the
  continuum of reduced dimensions.  Surprisingly, they are chiral, deeply
  related to fermion zero modes and anomalies in quantum
  chromodynamics. The chirality of the superfluid is manifested by a
  persistent anomalous mass current of atoms in the groundstate.  This
  spectacular quantum phenomenon is a prediction for future
  experiments.

\end{abstract}

\maketitle 


The $p$-wave Feshbach resonance has recently been demonstrated
accessible and robust in atomic gases of $^6$Li and
$^{40}$K~\cite{Regal+Jin:p_wave:03,Zhang+Salomon:04pre}.  Pioneer
theoretical
studies~\cite{Volovik:04,Ho+Diener:05,Gurarie:p_wave:04pre,Iskin+Melo:05pre}
point to  interesting new properties associated 
with the resonance.   
For a simple model, one may consider a system of spinless
fermionic atoms interacting through a $p$-wave potential,
isotropic in the 3D orbital space. Such an interaction would enjoy
the usual SO(3) spherical symmetry in orbital space. 
However, the $p$-wave
Feshbach resonances are split between the perpendicular and
parallel orbitals with respect to the magnetic field axis, due to the
(anisotropic) magnetic dipole-dipole interaction. This is known both
experimentally and
theoretically~\cite{Ticknor+:multiplet:04}.
The energy splitting, in units of Bohr magneton,  
is significantly greater than both the free atom Fermi
energy [for a typical gas density] and the resonance width [which is
{\it narrow}].  
Therefore,
it is  perhaps best to 
distinguish the $p_{x,y}$ orbitals  from the $p_z$
orbital and model the two
$p$-wave resonances separately.

\paragraph{The planar $p$-wave model}
Focus on the resonance involving the $p_{x,y}$
orbitals. I postulate a {\it planar} $p$-wave model for the atomic
Fermi gas with the interaction parametrized by a  single coupling
constant $g$. Using $a^\dag_\Bk$ and $a_\Bk$ as 
creation  and
annihilation operators for atoms with momentum $\Bk$,  the Hamiltonian reads   
\be
H= \sum_\Bk \eps_\Bk a^\dag_\Bk a_\Bk + {g\over 2 \SV}
\sum_{\Bq\Bk\Bk'} \vec{k}\cdot{\vec{k}^\prime} 
a^\dag_{\frac{\Bq}{2}+\Bk}a^\dag_{\frac{\Bq}{2}-\Bk}
a_{\frac{\Bq}{2}-\Bk'}a_{\frac{\Bq}{2}+\Bk'} \,, 
 \label{eq:H:pp}
\ee
where $\eps_\Bk=\frac{\Bk^2}{2m} -\mu$ is the energy spectrum 
of the free
atom, 
measured from the chemical
potential $\mu$, and $\SV$ is the space volume 
($\hbar\equiv1$ unless explicitly restored). A convention is adopted:  
the boldfaced vector labels three components $\Bk=(k^x,k^y,k^z)$ and
the arrow vector collects the $x,y$ components only, $\vec{k}=(k^x,k^y)$. 
When expanded in terms of the
spherical harmonics, the interaction term contains only $p_x$ and
$p_y$ orbitals as indicated by the arrow vector; it
explicitly breaks the 3D rotational invariance
down to a lower symmetry of
SO(2) that rotates about the $z$-axis.
Only $L_z$ of the three angular
momenta is conserved.  To compare with a physical system, 
the coupling constant $g$ is related to the $p$-wave scattering
length $a_1$ and effective range $r_0$ by $g\simeq \frac{24\pi \hbar^2 a_1
  r_0^2}{m}$, with $a_1<0$~\cite{Landau-Lifshitz:77QM:Sec133}.

As I shall show below, this atomic gas model predicts remarkable
properties, including a new domain wall soliton and an anomalous
quantum mass flow of atoms persisting in one direction along the wall.

\paragraph{Axial superfluid}

For $g<0$, the interaction term in Hamiltonian~(\ref{eq:H:pp}) can be
decoupled by introducing a planar vector field $\vec{\Phi}$ via 
the Hubbard-Stratonovich transformation, standard in the path integral
formalism of many-body
theory. 
The field operator $\vec{\Phi}_{\Bq}$ is  
identified by the quantum equation of motion with  
a composite, $p$-wave atom pair, 
\be
\vec{\Phi}_{\Bq} = -\frac{g}{\SV}\sum_\Bk \vec{k}\,  
a_{\frac{\Bq}{2}-\Bk}a_{\frac{\Bq}{2}+\Bk} \,.
\ee
$\vec{\Phi}$ is the
superfluid order parameter.
This order parameter is a complex, two-component vector in orbital
space. We parametrize it with four independent real variables as follows,
\be
\vec{\Phi}_0\equiv {\avg{\Phi^x_{\Bq=0}}\choose 
\avg{\Phi^y_{\Bq=0}}} 
= {\rho} e^{i\vartheta \Bid} e^{-i\varphi \sig_2} {\cos \chi
  \choose i \sin \chi} \,,
\ee
where $\Bid$ is a $2\times 2$ identity matrix and $\sig_2$ the
second Pauli matrix. 
In literature, one encounters other popular parameterizations,
such as writing the order parameter as a rank-2 tensor or a sum of 
real and imaginary vectors. Whatever other representations are
can be mapped, in a one-to-one correspondence manner, to the above
form. An advantage of our parametrization is that its symmetry transformation
property is made transparent.  The
matrices $\Bid$ and $\sig_2$ are the generators of the U(1) phase
and SO(2) orbital rotation transformations, respectively. So,
$\vartheta$ represents the phase of the atom-pair wavefunction and $\varphi$ is
the azimuthal angle (about the $z$-axis) of the  order parameter as a
{\it planar} 
vector in orbital space. 
For any finite modulus $\rho$, both symmetries are
spontaneously broken down.  
The third angle variable, $\chi$, transforms 
$\chi\rightarrow -\chi$ under the
time reversal symmetry.   

\begin{figure}[htbp]
\begin{center}
\includegraphics[width=0.97\linewidth]{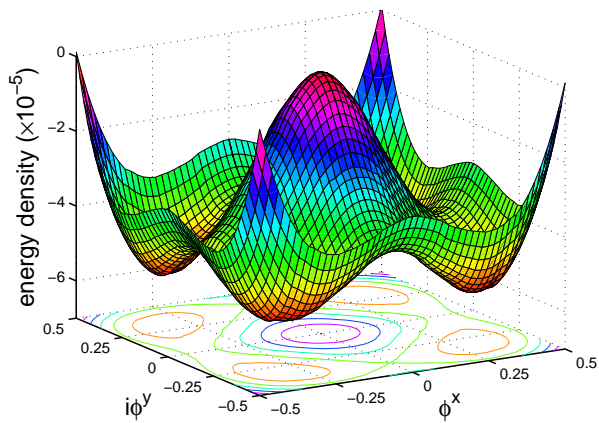}
\end{center}
\caption{Effective potential $V_\mathrm{eff}$ of Eq.~(\ref{eq:Veff}) 
per unit volume, with
coordinates
$[\phi^x, i\phi^y]=[\rho\cos\chi, -\rho\sin\chi]$ in order parameter
space and  $g/(2\pi)^3=0.5$.  
The energy and
length units are the Fermi energy (equal to $\mu$) and the Fermi
wavelength $k_F^{-1}$, respectively. 
}
\label{fig:eff_pot}
\end{figure}

For a homogeneous state of (spatially) uniform $\vec{\Phi}$, 
the effective potential (or free energy if
extended to finite temperature) density can be exactly calculated. I
found
\be 
V_\mathrm{eff} = \frac{\rho^2}{2|g|}-\int \frac{d^3\Bk}{2(2\pi)^3} \left[
  \sqrt{\eps^2_\Bk +|\vec{k}\cdot \langle{\vec{\Phi}_{0}}\rangle|^2} 
-|\eps_\Bk|\right] \,. \label{eq:Veff}
\ee
Notice that the potential does not depend on the phase $\vartheta$ and
the orbital angle $\varphi$ because of the symmetries.
  (Set $\vartheta=\varphi=0$
hereafter). 
Given a constant $g$, the effective 
potential is minimized
at a finite value $\rho=\rho_0$ and at $\chi=\pm {\pi\over 4},\pi\pm
{\pi\over 4}$ (see Fig.~\ref{fig:eff_pot} and Eq.~(\ref{eq:F[chi]})). 
$\rho_0$ is not universal, 
depending on the Fermi
and ultra-violet cutoff momenta.
The four minima of $\chi$ are degenerate.  A
$\pi$ phase identification, due to the U(1) phase symmetry of
the order parameter, reduces the four into two identical groups, each
of two minima.   So the angle variable
$\chi$ should be limited to half the scope of $2\pi$, say in 
$[-{\pi\over 2},{\pi\over 2}]$, to avoid the parametrization
redundancy. I emphasize that the two minima  
$\chi=+ {\pi\over 4}$ and $\chi=- {\pi\over 4}$ are discrete due to
time reversal symmetry.
This will be of
important topological implications later. 
Then, the (mean-field) states are 
the (axial) $p_x\pm ip_y$ 
superfluid:
$
\vec{\Phi}_0 
= {\rho_0\over\sqrt{2}} {1  \choose \pm i}
$
for $\chi=\pm{\pi\over 4}$, respectively. One of these two
states will become the groundstate, breaking 
the symmetry spontaneously.  
My finding agrees to the results of other different $p$-wave models
\cite{Ho+Diener:05,Gurarie:p_wave:04pre}.
The state carries a total angular momentum, estimated to be
$L_z^\mathrm{total}=  \pm {N_0\over 2}\hbar$ with $N_0$ the number of atoms in
the condensed pairs. $N_0$ is directly
accessible in current experiments.


\paragraph{Domain Walls}
The degeneracy of the $\chi=\pm {\pi\over 4}$ states motivates
us to look for a spatially nonuniform, soliton solution that
intervenes the two minima and varies smoothly from one to another 
in real space.  Such a configuration is
topologically stable since the two minima are related by  the
{\it discrete} time
reversal $Z_2$ symmetry, not by any continuous symmetries such as orbital
rotation and phase transformation.  

To uncover the soliton granted by the topology of broken time-reversal
symmetry, 
we consider temporally stationary,
spatially nonuniform configurations of $\chi(\Br)$
with $\Br$ the space coordinate while keeping the modulus
constant. The order parameter then becomes 
\be
\vec{\Phi}(\Br)\equiv {{\Phi^x(\Br)}\choose 
{\Phi^y(\Br)}} 
= {\rho_0} {\cos \chi(\Br)
  \choose i \sin \chi(\Br)} \label{eq:phi_dw}
\ee
where $\vec{\Phi}(\Br) =
\sum_\Bq e^{i\Bq\cdot \Br} \vec{\Phi}_\Bq$. 

Our goal is to find  an effective theory for the angle field 
$\chi(\Br)$
that retains both spatial derivative terms and
potential terms. Such an effective theory is obtained perturbatively 
through the
Wilsonian renormalization approach, by
integrating out
fermionic atoms at high energies (measured from the Fermi level) while keeping
low energy fermions within a thin shell
of thickness, say $\kappa$, along the Fermi sphere in
the momentum space.  
I have found the effective free energy functional
of $\chi(\Br)$, 
\be
F = \frac{\rho_0^2mk_F}{2C_\Phi} \int d^3 \Br \left[ 
  (\nabla\chi)^2 + \frac{1}{2\xi^2} \Big(1+
  \cos(4\chi)\Big)\right]\,. \label{eq:F[chi]}
\ee
The characteristic length ${\xi}$, emergent in the low energy 
effective theory, 
is the coherence scale on which the
order parameter varies. A perturbative
calculation determines ${\xi}= {k_F\over m\Del_0}\times 
C_\xi({\kappa\over k_F},{\Lambda\over  k_F}) 
$ 
where $\Del_0$ is the
maximum energy gap of quasiparticle excitations. 
$C_\Phi$ and $C_\xi$ are dimensionless {\it positive} constants depending
on the infrared and ultraviolet cutoffs ($\kappa$ and $\Lambda$), and
$k_F$ is the Fermi wavevector $k_F^2/(2m)=\mu$. [Detailed derivations,
including coefficients $C_\Phi$ and $C_\xi$,  
will be given 
elsewhere.]  From
either this perturbative calculation or the general
Ginzburg-Landau phenomenology, the order parameter should
vary slowly compared with the Fermi wavelength, i.e., $1/(\xi k_F) \ll 1$.
The effective theory is invariant under a  $Z_2$ time-reversal
symmetry, 
$\chi \rightarrow -\chi$.


The effective theory happens to be
a variant of  the well-known sine-Gordon Lagrangian  in the 3D Euclidean
space.  
It is known to possess (nonuniform) soliton
solutions~\cite{Rajaraman:87bk}. Applying the textbook method, we
find soliton and anti-soliton configurations (say, constant in $y,z$
directions),
\be
\chi_\pm(\Br) = \pm \arctan \Big(\tanh(x/{\xi})\Big)\,, \label{eq:chi=}
\ee
where `$\pm$' are the two topological charges 
(or winding numbers). For either configuration, $\chi(\Br)$ changes sign
when $x$ varies from 
$-\infty$ to $+\infty$, developing 
a wall that separates two domains of opposite angular
momentum. The domain wall 
is centered  at $x=0$, a position however set
arbitrarily for simplicity. 
The thickness of the
wall is characterized by $\xi$. 
Its energy per unit area is calculated to be 
$\rho^2_0/(C_\Phi\xi)$, proportional to $\Del_0/\xi^2$ (recall
$\Del_0$  the maximum energy gap).

\paragraph{Gapless chiral fermion bound states}

\begin{figure}[htp]
\begin{center}
\includegraphics[width=0.8\linewidth]{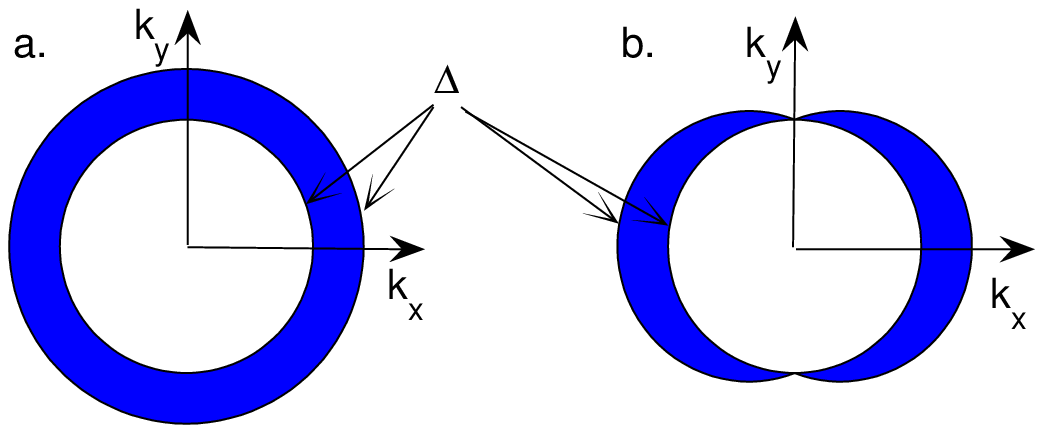}\\
\includegraphics[width=\linewidth]{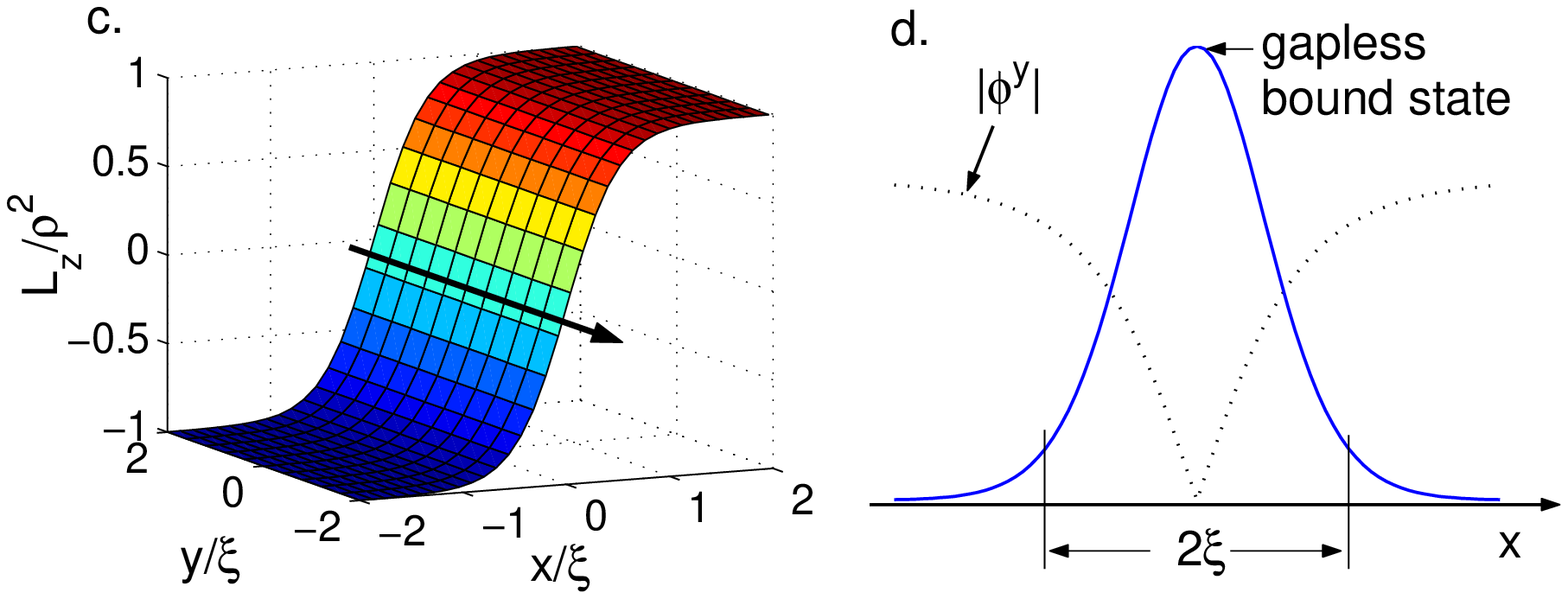}
\end{center}
\caption{Heuristic illustration of
gapless chiral quasiparticle bound states: (a) momentum-space 
energy gap in the axial
$p_x$+$ip_y$ superfluid state; (b) gap in the $p_x$ state; (c) the soliton
defect in the real space with the angular momentum 
$\vec{L}=-i{\vec \Phi}^* \times \vec{\Phi}=\hat{e}_z \rho_0^2 \sin\chi$; 
(d) the gapless bound state profile with 
$|\Phi^y|$ as the ``off-diagonal'' potential
(`$\cdots$').
The arrow in (c) indicates the direction of anomalous
mass flow.}
\label{fig:dwf}
\end{figure}

What happens to the quasiparticle states when the order parameter develops a
soliton defect?  Consider an order parameter configuration in real
space depicted in
Fig.~\ref{fig:dwf}c. 
$\chi$ varies
in a characteristic distance of
${\xi}$, far greater than the Fermi wavelength
$k_F^{-1}$. 
The fermionic
atoms view the soliton defect as a fairly flat, 
``classical'' off-diagonal potential that
hybridizes the particle and hole states of atoms.
So we can loosely speak of three different regions 
along the $x$-axis---left ($p_x-ip_y$), domain wall ($p_x$ state), and
right ($p_x+ip_y$)
boxes. 
Now imagine a momentum space for each `macroscopic 
box', each beginning with 
free fermionic atoms labeled by $3$-component definite
momentum $\Bk$.
An important observation is that the off-diagonal pairing potential
seen by the atomic states in the $k_y$-axis direction
must be vanishingly small in the domain wall box
(exact zero at the center of the box) but maximizes at infinities 
in the left and right boxes. 
Therefore, like the Caroli-de Gennes-Matricon
state~\cite{Caroli+deGennes:64}  
in the vortex core of a superconductor where the order parameter
vanishes, we expect to see zero energy states peaked at the domain
wall  and bound by the rising $p_y$-wave component of the 
gap potential at far away from the wall. Unlike the 
vortex core states, the domain wall fermions are only bound in the
direction perpendicular to the wall ($x$-direction) and disperses like
in continuum in the parallel directions (Fig.~2).

Such heuristic argument can be made rigorous by analyzing the
Hamiltonian~(\ref{eq:H:pp}) in the axial state with the domain
wall defect. 
The (mean-field) Hamiltonian can be diagonalized
by using the Bogoliubov transformation 
\be
a(\Br) = \sum_{n\Bk_\parallel} \left[ u_{n\Bk_\parallel}(\Br) \,
c_{n\Bk_\parallel} + v^*_{n,-\Bk_\parallel}(\Br) \,
c^\dag_{n,-\Bk_\parallel}  \right]\,
\ee
where  $\Bk_\parallel=(0,k_y,k_z)$ is  a  
momentum  parallel to the domain wall, 
and $n$ a quantum number
from the quantization of $x$-direction. 
The unitarity of the transformation requires that 
$\int d^3\Br [u_{n\Bk_\parallel}(\Br) 
u^*_{n^\prime\Bk^\prime_\parallel}(\Br) + 
v_{n\Bk_\parallel}(\Br)v^*_{n^\prime\Bk^\prime_\parallel}(\Br)] = 
\del_{nn^\prime}\del_{\Bk_\parallel,\Bk^\prime_\parallel}$\,. 
The unbroken symmetries of translation
parallel to the domain wall simplify the eigenstates, 
$$
 \Big(u_{n\Bk_\parallel}(\Br)\,,
v_{n\Bk_\parallel}(\Br)\Big)= \Big(\tilde{u}_{n\Bk_\parallel}(x)\,, 
\tilde{v}_{n\Bk_\parallel}(x)\Big) {e^{i\Bk_\parallel\cdot \Br} \over 2\pi}\,.
$$
 
We are primarily interested in finding the low energy quasiparticle
states 
that have 
a parallel momentum in order of $|\Bk_\parallel|\sim k_F$ 
(Fermi wavevector) and vary slowly
in $x$ direction in real space. 
For 
$\xi k_F \gg 1$,  the eigenstates are dictated by the following
Bogoliubov-de Gennes  equations
$$
\begin{array}{c}
-i\rho_0 \big(\cos\chi \partial_x - k_y \sin\chi\big)
\tilde{v}_{n\Bk_\parallel} = (E_{n\Bk_\parallel}- \eps_{\Bk_\parallel})
\tilde{u}_{n\Bk_\parallel} \,, \\
-i\rho_0 \big(\cos\chi \partial_x + k_y \sin\chi \big) 
\tilde{u}_{n\Bk_\parallel} = (E_{n\Bk_\parallel} + \eps_{\Bk_\parallel})
\tilde{v}_{n\Bk_\parallel}\,,
\end{array}
$$
where $\chi= \chi_+(\Br)$ [Eq.~(\ref{eq:chi=})] and $\rho_0$
is a constant related to the maximum gap by $\rho_0k_F =\Del_0$. 

The above equations have  symmetries. While reversing the
sign of both $\chi$ and $k_y$ simultaneously or reversing the sign of
$k_z$ independently, the eigenvalues are invariant. One can also
prove that given any  
eigenstate
$(\tilde{u}_{n\Bk_\parallel},\tilde{v}_{n\Bk_\parallel})$ with
eigenvalue $E_{n\Bk_\parallel}$, there always exists another
eigenstate,
$(-\tilde{v}_{n\bar{\Bk}_\parallel},\tilde{u}_{n\bar{\Bk}_\parallel})$, 
with an eigenvalue of opposite sign, $-E_{n\bar{\Bk}_\parallel}$,
where  $\bar{\Bk}_\parallel=(0, -k_y,k_z)$. 
Such a doubling of eigenstates are not surprising in a
superfluid system that had originally enjoyed time-reversal and parity
symmetries before undergoing superfluid.  
However, as I shall show later, 
the phenomenon of doubling eigenstates breaks down when chiral
fermions appear. There, the physical eigenstates are no longer symmetrically
available upon reversing the sign of momentum, say $k_y\rightarrow
-k_y$.

The reduced 1D eigenvalue problem resembles, but differs in
fundamental ways from,  the Dirac equation
of one spatial dimension analyzed by Jackiew and
Rebbi~\cite{Jackiw-Rebbi:76} who found the chiral zero-energy bound
state (zero fermion mode). In our case, we would expect 
that their zero-energy mode translates into gapless chiral 
quasiparticle  bound
states (chiral fermions).  
Indeed, focusing on the gapless branch ($n=0$), I found two 
branches of such states. They are
\be
\begin{array}{ll}
E^+_{0\Bk_\parallel}=+\eps_{\Bk_\parallel}:  &
\tilde{u}^+_{0\Bk_\parallel} \simeq \left[\cosh({x/
  {\xi}})\right]^{-k_y{\xi}}, \,
\tilde{v}^+_{0\Bk_\parallel} =0\,,\\
E^-_{0\Bk_\parallel}=-\eps_{\Bk_\parallel}: &
\tilde{u}^-_{0\Bk_\parallel} =0, \,
\tilde{v}^-_{0\Bk_\parallel} \simeq [\cosh({x/
  {\xi}})]^{k_y{\xi}}
\,,
\end{array} \label{eq:bound_state}
\ee
with `$\simeq$' meaning a proper normalization to be achieved yet. Clearly, the
$E^+$ and $E^-$ branches are only normalizable (so  become 
physical bound states) for 
$k_y\geq0$ and  $k_y\leq 0$, respectively. The superscript `$\pm$' is best
interpreted as a sign of {\it positive} or {\it negative}
chirality for the existing gapless fermion modes. 
For either
chirality the energy $E_{n\Bk_\parallel}$ can be positive or negative,
depending on the value of $\eps_{\Bk_\parallel}=\Bk_\parallel^2/(2m)-\mu$. 
The doubling of eigenvalues is removed by the presence of
chirality, with only one mode being physical bound state, supporting
the claim made early. 
For gapped
branches (denoted with $n>0$), 
I would speculate, based on the study of domain wall
quasiparticle excitations in $^3$He~\cite{Ho+Wilczek:84},  
that bound states exist for both
positive and negative momentum $k_y$ (hence not chiral) 
and the doubling should restore.

\paragraph{Anomalous quantum mass flow}



With the eigenfunctions diagonalizing the superfluid Hamiltonian, we
can directly calculate the atom occupation number in the state vector
space, specified by the quantum numbers
$|n\Bk_\parallel\rangle$. I found that the atom occupation number per
state is
$
N_{n\Bk_\parallel}=\int dx\,  |\tilde{v}_{n\Bk_\parallel}(x)|^2
$,
derived from  the total atom number
$
N= \int d^3 \Br \avg{a^\dag(\Br) a(\Br)} = \sum_{n\Bk_\parallel}
N_{n\Bk_\parallel}\,.
$
Clearly, the discrete quantum number $n$
takes the role of $k_x$ for a homogeneous superfluid state (where 
the domain wall defect is absent). 
Finite temperature effects, which are not included in this
paper, will modify the above result through extra terms and factors 
dependent on the Fermi-Dirac distribution function. The occupation
weight is essentially given by the eigenfunction $\tilde{v}$ alone,
consistent with the result established for the homogeneous
superconductivity. 
The physical normalizability
constrains that
non-zero $\tilde{v}_{0\Bk_\parallel}$ exist only for negative
chirality of $k_y\leq 0$ (Eq.~(\ref{eq:bound_state})). We then must conclude that 
the atoms of
quantum number $n=0$ only occupy
half the reduced 2D $k_y$-$k_z$ momentum plane (parallel to the domain
wall), with a 
spectral weight $|\td{v}_{0\Bk_\parallel}|^2$.
The another half space is exactly empty  for the gapless states
in the superfluid groundstate (zero temperature). Bear in mind that the
$n> 0$ states are different.
This is the origin of an anomalous chiral mass flow of atoms.   

A direct calculation of the
mass current of atoms verifies the above intuitive speculation. 
I found that
the chiral gapless bound states give rise to  
a spectacular anomalous current of atoms
equal to
$$
\Bj = -{\hbar k_F^3\over 6\pi^2} \hat{e}_y \,, \quad \mbox{($\hbar$
  restored for clarity)}
$$
per unit length in $z$-axis (perpendicular to the flow). The
current persists without an additional {\it external}  field. Its
direction is set by the topological charge of the soliton
defect. For an anti-soliton, the anomalous mass flow reverses
direction. In real experiments that the magnetic field sets the
$z$-axis $(\BH\parallel \hat{e}_z)$, the prediction is that 
the anomalous mass current  flows parallel to the domain wall in the
direction of   
$\nabla L_z\times \hat{e}_z$ where $\nabla L_z$ is taken as the
gradient of the angular momentum $L_z$ at the wall (Fig.~2c). 
The mass flow returns at the boundaries of the atomic trap which constitute
an anti-soliton (anti-domain wall).  The total mass current, therefore, does
not violate conservation law. 

At finite temperature, excitations with positive energy will
proliferate those unoccupied states in the empty half space and/or
deplete the occupied states.
The finite temperature effect is then to create a counter
mass flow in opposite direction.  I conjecture that the
anomalous mass flow should decrease with increasing temperature and
should be strongest in magnitude for temperatures
(times the Boltzmann constant)
below the level splitting of quasiparticle bound states, characteristically
given by $\sim \Del_0^2/\mu$.  

\paragraph{Discussion}
The anomalous current is reminiscent of that in the $^3$He
A-phase~\cite{Ho+Wilczek:84,Balatskii:86,Stone:87} and in the PbTe
semiconductor~\cite{Fradkin_parity:86}, but differs in nature.  First,
the domain wall defect I discovered is a new kind different than the
``twist'' texture of Ref.~\cite{Ho+Wilczek:84,Balatskii:86,Stone:87}
for $^3$He-A. In the latter, the angular momentum $\vec{L}$ sweeps its
direction in real space with a unit magnitude. This domain wall
is however that $\vec{L}$ is fixed in the $\hat{z}$-axis direction but
changes sign and magnitude across the wall (Fig.~\ref{fig:dwf}c).
Second, treating the pairing potential semi-classically, the
quasiparticles see a gapless line (in the $k_y$-$k_z$ plane of
Fig.~\ref{fig:dwf}b) in the domain wall region in this case as opposed
to nodal points everywhere in the $^3$He-A case.  These features,
among others, are new for the atomic gas of {\it anisotropic} $p$-wave
Feshbach resonances. The $^3$He-A
results~\cite{Ho+Wilczek:84,Balatskii:86,Stone:87} do not directly apply.

This domain wall defect seems interesting and new for
the experiment of atomic Fermi gases, currently  developing rapidly.
I speculate that superfluid domains of opposite
angular momentum may be easier to realize than a homogeneous (axial)
superfluid, since the latter requires rotating the whole gas (or
other means) for a net input of macroscopic
angular momentum.
I am not aware whether the anomalous current predicted for the $^3$He-A
has been observed experimentally. It would be 
first  if the $p$-wave resonant atomic gas shows this spectacular
quantum anomaly. That would  be also  of interest to the study
of lattice quantum
chromodynamics attempting to simulate chiral quark fields by
fermion zero-energy mode bound at a domain wall~\cite{Kaplan:1992bt}.

\bibliography{anomalous_flow}
\end{document}